\documentstyle{ioplppt}
\begin{document}
\jl{1}
\letter{ Measurement, Decoherence and Chaos in Quantum Pinball}
\author{C. Dewdney\dag, Z Malik\dag}
\address{\dag\ Physics Division, University of
Portsmouth, Portsmouth PO1 2DZ, United Kingdom}

\begin{abstract}
The effect of introducing measuring devices in a ``quantum pinball'' system is
shown to lead to a chaotic evolution for the particle position as defined in
Bohm's approach to Quantum Mechanics.
\end{abstract}

\section{Introduction}
One of the major problems in the study of the quantum mechanical behaviour
of
classically chaotic systems, in the context of the orthodox interpretation, is
the
inapplicability of the usual  means of describing the dynamics of any system in
terms
of well-defined trajectories. In contrast to the orthodox approach to quantum
mechanics Bohm's approach\cite{deB26,deBbook,Bohm52} does allow the
description of quantum systems in terms of well-defined trajectories and so
overcomes this particular limitation. The Bohm trajectories are derived from
the wavefunction according to the guidance condition
\begin{equation}
\vec{p}=\Im\left [\frac{\Psi^\dagger\nabla\Psi}{\Psi^\dagger\Psi}\right ]
\end{equation}
where $\vec{p}$ is the momentum and we assume $\hbar=1$.
In a recent work\cite{MD95,zmthesis} we used Bohm's approach to explore the
quantum mechanical behaviour of the (classically chaotic) kicked rotor and,
notwithstanding the nonlinear quantum potential which arises in this
description, found that neither an isolated rotor in a superposition of states
nor
an otherwise isolated but periodically kicked rotor showed any evidence of
chaotic behaviour.
The Schr\"{o}dinger equation is non-mixing (two initially almost identical
wavefunctions remain almost identical under the Schr\"{o}dinger evolution)
and this coupled with the fact that Bohm trajectories do not cross in the
configuration - space of the system,
which in this is case is one dimensional, makes the conclusion that chaos is
not observed not surprising.

Thus far there is little in the extensive quantum chaos literature concerning
the
effects of measurement on a quantum system. This is probably because, from
the orthodox point of view, the effect of measurement on a quantum system is
to introduce an inherent randomness of outcomes and this essentially removes
the system from the arena of interest for quantum chaology; chaos should arise
from the internal dynamics of the system, not as a result of an external
randomising influence.
The introduction of measuring devices, when they are simply considered as
other quantum systems, enlarges the configuration space of the (now
compound) system. Their principle function is to introduce bifurcations in the
evolution of the probability density in configuration space which then flows
into separate regions. A good measuring device also ensures that the
possibility
for interference between these different regions in the future is negligable.
In
common text-book parlance the configuration-space wave then collapses (by
some unspecified non-quantum mechanical process) in a random manner into
just one of these regions.

In contrast to the usual description,  Bohm's approach has no need for
wavepacket collapse. There is a well-defined outcome in each individual
measurement since the actual configuration of the compound system is given by
a well defined point in configuration space (just as in classical mechanics)
and,
according to the equations of motion, this point moves into just one the
separate
regions that develop as the interaction proceeds. Thus a unique correlated
state
of the measuring device and the measured system is selected. If the device is a
good one then, for the practical purposes of calculating the future behaviour
of
the system, one may neglect all the unoccupied regions of the configuration
space and just employ that component of the superposition which is effective in
determing the future behaviour. Wave packet collapse to an effective
wavefunction is, in Bohm's theory, just a calculational convenience that one
may employ.

 At the end of  our above-cited paper on the kicked rotor we suggested that, if
the rotor were
to be subjected to repeated measurements of the angular momentum, one after
each kick,
then the evolution of the effective wavefunctions associated with different
initial
conditions could indeed be divergent, leading to chaotic behaviour for the
motion of
the rotor. For the case of the rotor\cite{MD93} we have shown that the outcome
of a
measurement of angular momentum is dependent not only on the initial state of
the rotor but also on the initial coordinates of the rotor's centre of mass
within
its wavepacket as it enters the Stern-Gerlach apparatus. After the rotor
suffers a
kick and enters a superposition of angular momentum eigenstates a subsequent
measurement will lead to the emergence of a  number of wave packets from the
exit from the Stern-Gerlach field and, given the fact that the Bohm
trajectories
may not cross, there will clearly be a corresponding number of bifurcations in
the associated trajectories. Each one of these emerging wavepackets is
associated with a different angular momentum component eigenstate and since
the actual rotor position must, in any
individual case, enter just one of the packets as they separate in space, the
effective
wavefunction (the one which determines the rotor's behaviour) becomes just
one of
the eigenstates. In Bohm's theory we are justified in ignoring the packets not
containing the actual rotor centre of mass coordinate for the purposes of
calculation providing only that the packets associated with different outcomes
remain orthogonal either in virtue of spatial separation or the functioning of
some complex recording device with many degrees of freedom. That is, in
modern parlance, the alternatives must decohere. On receiving the next kick
the effective angular momentum eigenstate wavefunction becomes a
superposition of angular momentum eigenstates once more. The values of the
coefficients of the eigenstates in this superposition will depend on the
rotor's
state before the kick. Consequently the sequences of effective wavefunctions,
generated by rotors with arbitrarily close initial positions, in a sequence of
kicks followed by angular momentum measurements will in general diverge.

The picture is this: two identical rotors, differing only slightly in the
values of
their initial positions within their identical centre-of-mass wave-packets are
subjected to a series of kicks. In the absence of any other interactions the
behaviour of the two rotors will remain closely correlated. The situation is
radically different if, between kicks, an angular momentum measurement is
carried out. Initially, for a few kick and measurement pairs,  the trajectories
of
the centre of mass of each rotor will be closely correlated and the two
sequences
of eigenstates resulting from the measurements will be identical. Eventually
the
center of mass coordinates of the two rotors will be placed either side of one
of
the bifurcation points that arise on measurement. Then the effective eigenstate
for each rotor will be different after the measurement. On the next kick the
expansion coeficients of the angular momentum eigenstates will be different
and the behaviour of the rotors will no longer be correlated.

\section{Quantum Pinball}
In order to explore the type of behaviour which arises with a kicked and
measured rotor, but in the context of a simple system for which
the outcome of the measuring interaction is merely twofold, we consider here a
``quantum-pinball system''. The model-system we have in mind is constructed
from a series of potential barriers, each with a transmission coefficient of
one-
half, which are arranged in a typical fairground pinball array (essentially on
a
triangular lattice within a triangle). A wave packet incident on the first
barrier
at the apex of the set of pins splits into two equal-sized packets which then
propagate on towards two
further barriers where the splitting behaviour is repeated. The significant
coordinate is clearly perpendicular tothe barriers, the motion parallel is
unaffected by the
barriers.

We shall consider two variants of the pinball, the first is as described so
far, but
the
second has the addition of  measuring devices, capable of recording the passage
of the
particle, in each arm of the network. Before describing the motion of the
particle
through the whole pinball let us examine the behaviour of a particle as its
wavepacket scatters from a single potential barrier.

{}From the fact that the Bohm-trajectories may not cross we can deduce that all
those particles which approach the barrier within the trailing half of the
packet
must be reflected, whilst those in the front half of the packet must be
transmitted.\cite{DH82} Thus this process, widely held to be inherently
random in the usual approach to quantum mechanics, can be given a fully
deterministic description in Bohm's theory. On scattering from the barrier the
wavefunction develops into a superposition
\begin{equation}
\Psi_{\mbox{inc}}=\Psi_{\mbox{ref}}+\Psi_{\mbox{trans}}
\end{equation}
but in each individual case there is a definite outcome as the Bohm trajectory
must lead into just one of the packets. Whilst the two packets remain
orthogonal, in virtue of their separation in space, only that part of the
superposition associated with the actual particle position is active in
determining the behaviour of the particle (this we refer to as the effective
wavefunction). If the two packets were to be recombined, as happens in the
second level of the pinball, then the reflected and transmitted packets would
interfere and both would be relevant in determining the particle's behaviour.

The situation is very different if measuring devices, capable of recording the
passage of the particle, are introduced in the transmitted and reflected paths.
The wave function develops in the following way
\begin{equation}
\Psi_{\mbox{inc}}\Phi_0(1)\Phi_0(2) \rightarrow
\Psi_{\mbox{ref}}\Phi_1(1)\Phi_0(2)+\Psi_{\mbox{trans}}\Phi_0(1)\Phi_1(2)
\end{equation}
where $\Phi_1$ and $\Phi_0 $ are orthogonal (and non-overlapping in the
associated coordinate space) and the number in parenthesis labels the detector.
Under these circumstances if the two beams emerging from the beam splitter
are recombined they do not interfere and the behaviour of a particle associated
with one part of the superposition remains independent of the other part of the
superposition. (The alternatives decohere.) In the configuration space spanned
by the particle coordinates and by all the measuring  device coordinates the
Bohm trajectories do not cross, but in the subspace spanned by just the
particle
coordinates the trajectories may cross.

With this in mind let us now consider the quantum pinball. In the absence of
the measuring devices the incident wavepacket splits at each barrier and
subsequently interferes at each further barrier with the packets that have
propagated along the various alternate paths to that barrier. Given that the
Bohm trajectories in this two-dimensional system cannot cross the order of the
trajectories in the initial packet is maintained as the packet propagates
through
the system and two trajectories that start close together must remain close
together.
Two such quantum trajectories are shown in figure 1.

The particle trajectories are very different if we include the measuring
devices.
The splitting behaviour at the first barrier remains the same as in the case
discussed above, but subsequently the system evolves in a very different
manner. The presence of the measuring devices ensures that there is no
interference between any of the packets that propagate along different paths in
the pinball apparatus. These alternative paths may then be said to form a set
of
consistent histories. Consequently the behaviour of the trajectories at each
barrier is the same as their behaviour at the first barrier: the packet splits
at the
centre, those in front are transmitted whilst those behind are reflected. Which
path a given particle takes through the pinball is determined at the outset by
its
position within the initial packet. Every time the particle scatters from a
barrier
its position with respect to the centre of the emerging packet is different. In
fact
a simple calculation shows that the sequence of positions of the particle in
the
wave packet, as the particle scatters from successive barriers, is given by
\begin{equation}
x_{n+1}=2x_n\bmod 1
\label{Bern}
\end{equation}
This behaviour of the particle coordinate is identical to that discussed by
Goldstein, D\"{u}rr and Zanghi \cite{Goldstein} in the context of a particle
oscillating back and forth in a double-well potential and subjected  to
repeated
position measurements. In our example of chaotic Bohm motion, the position of
the particle relative to the centre of the wave packet, as it scatters from the
barriers, determines the path taken through the pinball system. We can think of
the particle position in the wavepacket as an internal coordinate and the
centre
of the packet actually containing the particle as an external or macroscopic
coordinate. Since the internal coordinate behaves chaotically according to
equation~(\ref{Bern}) the external coordinate follows an apparently random walk
through the pinball. If the system is set up twice, with slightly different
initial
internal coordinate in each case, the corresponding motions of the wave packets
through the pinball soon become totally uncorrelated, as the iterations of
equation~(\ref{Bern}) quickly diverge. Two such trajectories are shown in
figure
2. In practise of course it is not experimentally possible to set the system up
with the same wave packet but different initial particle position within the
packet, any attempt to determine the particle position will also alter its wave
function.

When the particle in the pinball is not measured we see that the Bohm
trajectories are very different to those of a classical particle in such a
pinball
device. With the measuring devices in place the trajectories are similar to
those
of a classical particle. In the WKB limit (in which the pins are smoothly
varying  potentials in the distance of one wavelength) and when the wavepacket
is small compared to the size of the pins one would expect the whole
wavepacket to follow a single path through the pinball. In this case the Bohm
trajectories would not diverge, they would all be grouped around the same path.
As the packet spreads it would begin to split on scatterring from the pins. If
no
measuring devices are present on the paths the wave-like behaviour discussed
above will develop, with the measuring devices in place the internal coordinate
will become chaotic and the packet trajectory will become random. However, if
the measuring devices also reconstitute the original packet dimensions on each
occasion (thus preventing the spreading of the packet from allowing different
paths through the pinball developing) the packet will continue to follow a
classical path, as will the Bohm trajectories.
\section{Conclusion}
Since measurements are held to be inherently random processes in orthodox
interpretations of quantum theory the study of their effects generally falls
outside the arena of interest for quantum chaologists. In Bohm's approach
measurements are deterministic dynamical processes and their effect on the
evolution of the coordinates of a system is a legitimate case for study in the
search for quantum chaos. Here, in accordance with the results of D\"{u}rr
et.al., we have shown that quantum chaos will arise in simple quantum
systems, such as the kicked-rotor or pinball systems, when they are subjected
to
repeated measurements.

The quantum kicked-rotor when completely isolated does not show chaotic
motion, so how does chaos arise in the classical limit? One answer is that
classical and hence chaotic behaviour in the motion of kicked rotor will
develop
when it is allowed to interact with certain types of environment (here a set of
measuring devices, which may themselves mimick a more complicated
interaction with an more general environment capable of storing information
about the rotor's state) in such a way that the various alternatives that arise
in
the time-development of the system's wave function decohere.

\ack
The authors wish to thank Shelly Goldstein for pointing out his reference
\cite{Goldstein} and for useful e-discussions.

\Figures
\begin{figure}
\caption{Two Bohm trajectories in the quantum pinball with slightly different
initial internal coordinates, in the absence of measuring devices. There is no
divergence.}
\end{figure}
\begin{figure}
\caption{Two Bohm trajectories in the quantum pinball started with slightly
different
initial internal coordinates,  with the meaasuring devices in place. The
trajectories diverge as the internal coordinate executes completely chaotic
motion.}
\end{figure}

\begin{thebibliography}{99}

\bibitem{deB26} de Broglie L (1926) {\it C.R.  Acad.  Sci.  Paris.}  {\bf 183}
447;
{\bf 185} 580 (1927)
\bibitem{deBbook}\dash (1960) {\it Nonlinear  Wave  Mechanics}  (Amsterdam:
Elsevier)
\bibitem{Bohm52} Bohm D (1952) {\it Phys. Rev.} {\bf 85} 166, 180
\bibitem{bohm51}\dash (1951) {\it Quantum Theory} (New York: Prentice Hall)
\bibitem{MD95} Malik Z, Dewdney C (1994) submitted to {\it Phys. Lett. A}
\bibitem{zmthesis} Malik Z (Nov 1994) {\it PhD Thesis} (Univ. of Portsmouth)
\bibitem{MD93} Dewdney C, Malik Z (1993) {\it Phys. Rev. A} {\bf 48}, pp
3513--24
\bibitem{DH82} Dewdney C, Hiley B J (1982){\it Foundations of Physics} {\bf 12}
27
\bibitem{Goldstein} D\"{u}rr D, Goldstein S, Zanghi N (1992) {\it Jour. Stat.
Phys}
{\bf 68}
\end{thebibliography}
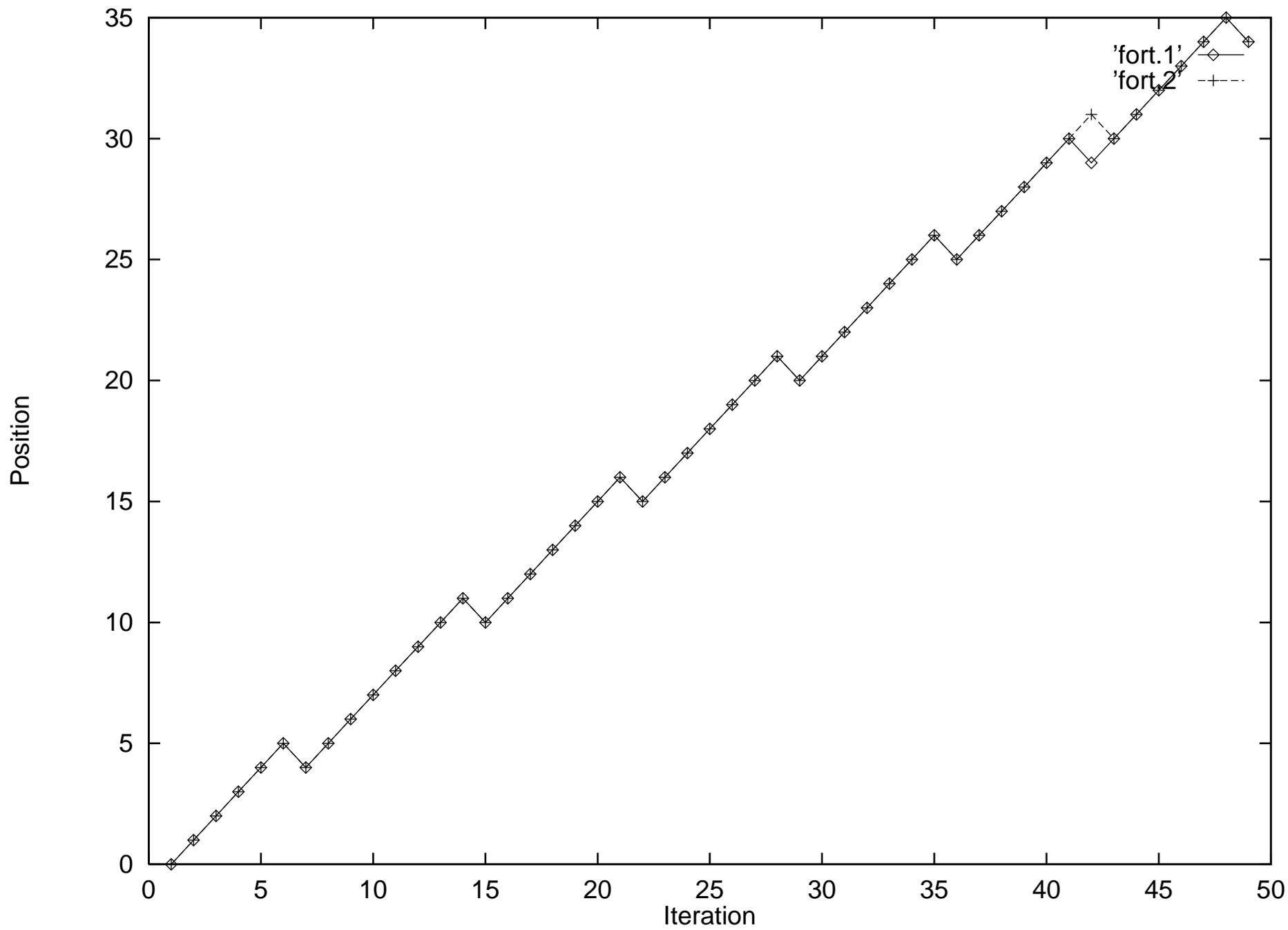
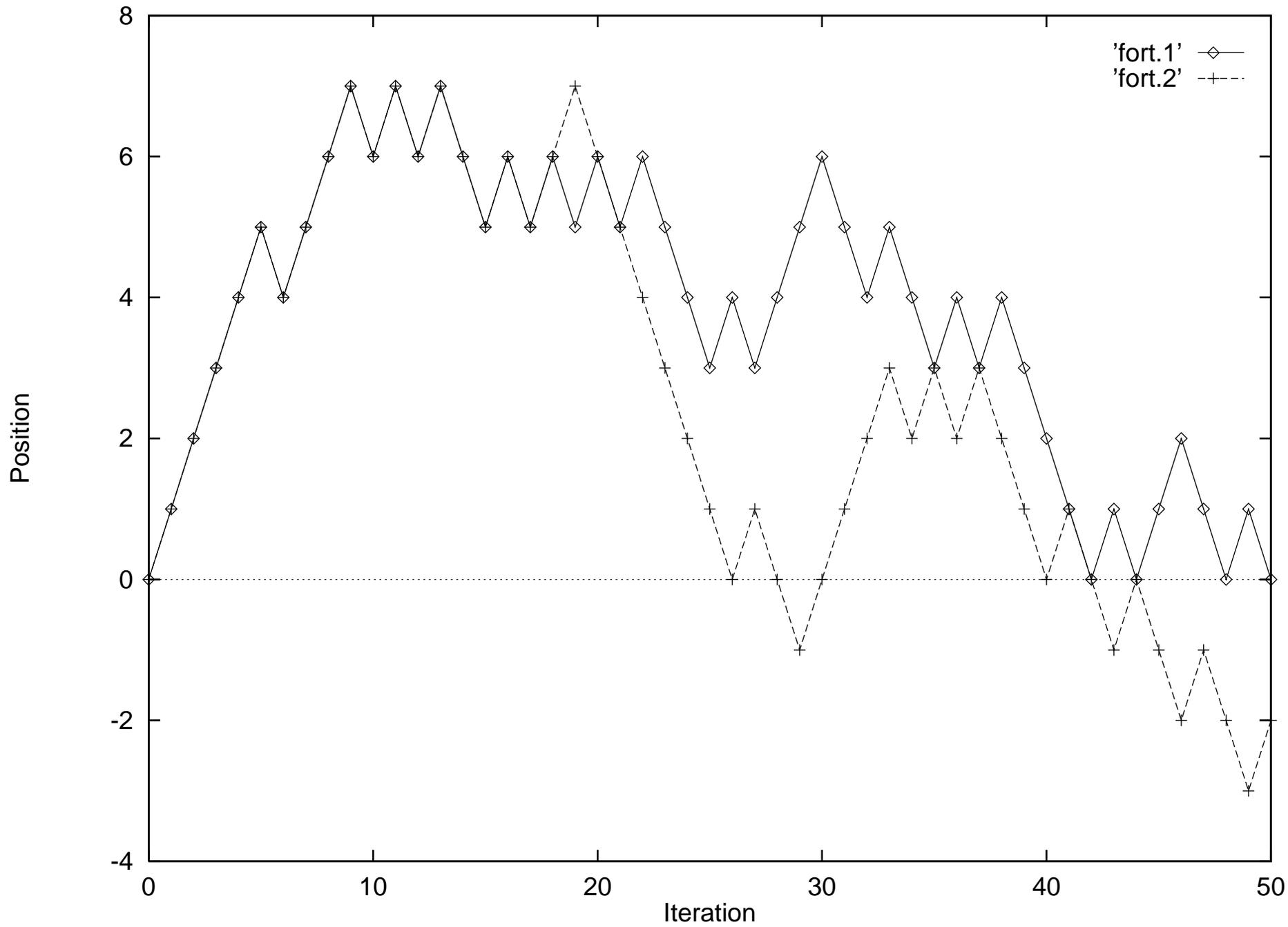
\end{document}